\documentclass[showpacs,twocolumn,preprintnumbers,amsmath,amssymb]{revtex4}
\usepackage{graphicx} 

\begin{document}

\title{\bf Heat Transport through Rough Channels}

\author{J. S. Andrade Jr.}
\email{soares@fisica.ufc.br}
\affiliation{Departamento de F\'\i sica \& Programa de P\'os-Gradua\c c\~ ao 
em Engenharia Qu\'\i mica,\\
Universidade Federal do Cear\'a, 60451-970 Fortaleza, Cear\'a,
Brazil}

\author{E. A. Henrique}
\affiliation{Departamento de F\'\i sica, Universidade Federal do Cear\'a,\\ 
60451-970 Fortaleza, Cear\'a, Brazil}

\author{M. P. Almeida}
\affiliation{Departamento de F\'\i sica, Universidade Federal do Cear\'a,\\ 
60451-970 Fortaleza, Cear\'a, Brazil}

\author{M. H. A. S. Costa}
\affiliation{Departamento de F\'\i sica, Universidade Federal do Cear\'a,\\ 
60451-970 Fortaleza, Cear\'a, Brazil}

\date{\today}

\begin{abstract}
We investigate the two-dimensional transport of heat through viscous
flow between two parallel rough interfaces with a given fractal
geometry. The flow and heat transport equations are solved through
direct numerical simulations, and for different conduction-convection
conditions. Compared with the behavior of a channel with smooth
interfaces, the results for the rough channel at low and moderate
values of the P\'eclet number indicate that the effect of roughness is
almost negligible on the efficiency of the heat transport system. This
is explained here in terms of the Makarov's theorem, using the notion of
active zone in Laplacian transport. At sufficiently high P\'eclet
numbers, where convection becomes the dominant mechanism of heat
transport, the role of the interface roughness is to generally
increase both the heat flux across the wall as well as the active
length of heat exchange, when compared with the smooth channel. 
Finally, we show that this last behavior is closely related with
the presence of recirculation zones in the reentrant regions of the
fractal geometry.
\end{abstract}


\maketitle

\section{Introduction}

The study of transport phenomena in irregular media is fully
recognized nowadays as an important research subject with many
technological and industrial applications. In particular, the role of
the morphology of an active interface on the efficiency of heat and
mass transport devices has also been a theme of great interest for the
scientific community. For example, the development of a modern
approach for the project of heat exchangers must necessarily take the
detailed description of the surface geometry as a possibility for the
proper design of these devices.  It is well-known in this field that
the inclusion of extended surfaces or fins to the surface of the heat
exchanger can substantially increase the rate of heat transfer in the
system and, therefore, the efficiency of the equipment
\cite{Arpaci66,Incropera90}.

For all practical purposes, the restrictions imposed to the
geometrical features of an exchange interface should be the result of
an intimate balance between its complexity and cost from one side, and
the performance of the equipment from the other. In any case, an
important issue in the design of heat exchangers is certainly the
extent to which the heat transfer capacity of the system is dictated
by the available surface area. For instance, under diffusion-limited
conditions, the access of heat to an irregular surface can be highly
nonuniform due to {\it screening} effects. In this situation, the
regions corresponding to fins or extended protrusions will display a
higher activity as opposed to the deep parts of fjords which are more
difficult to access. As a result, under strong diffusive constraints,
the efficiency of the surface for heat exchange can be substantially
different from the one expected, considering its total area, the
intrinsic conductive properties of the material, and the transport
properties of the heating (or cooling) fluid.
 
An analogous problem can be found in the context of electrochemistry.
The nonuniform activity during the linear transport through irregular
interfaces of electrodes has been explained in terms of a screening
effect which is entirely similar to the one observed in heat
transfer. The extensive research developed on this subject in recent
years \cite{Sapoval94,Pfeifer95,Sapoval96} has been mainly devoted to
the introduction, calculation and application of the concept of {\it
active zone} in the Laplacian transport to and across irregular
interfaces. For example, through the coarse-graining method proposed in
\cite{Sapoval94} it is possible to compute the flux through electrode
surfaces from their geometry alone, without solving the general
Laplace problem. In a subsequent study \cite{Sapoval01}, it has been
shown that this technique provides consistent predictions for the
activity of irregular catalyst surfaces. More recently, the activity
of irregular absorbing interfaces operating under diffusion-limited
conditions has been investigated through nonequilibrium molecular
dynamics \cite{Andrade03}. The simulation results indicate that the
extent of the interface that is significantly active is rather
sensitive to the governing mechanism of transport. Precisely, the
length of the active zone decreases continuously with density from 
the Knudsen to the molecular diffusion regime.

The study of laminar-forced-convective heat transfer through tubes 
or between parallel plates with smooth walls is well known in the
literature as the {\it Graetz problem} \cite{Graetz}. Graetz solved the
problem analytically with the assumptions of steady, irrotational 
and incompressible flow, constant fluid properties, fully developed
velocity profiles, and in the absence of energy dissipation effects.
Much effort has been dedicated in the past years to improve 
the accuracy of the Graetz solution and generalize the Graetz
problem to include other types of boundary conditions
\cite{Randall97,Dennis56,Coelho01,Telles02,Sellars56,Brown60,Mikhailov97}.  
The development and solution of more realistic models for heat
transport in internal flow configurations and under forced convection
conditions has also been a research theme of great scientific interest
\cite{Johnston94,Papoutsakis81,Lahjomri03}. Few studies, 
however, have been dedicated to the investigation of an important
extension of the Graetz problem, namely, the case in which the flow
and heat transfer processes are confined to walls of irregular
geometry.

In the present work we apply the concept of active zone to investigate
the steady-state transport of heat in a fluid flowing through a
two-dimensional channel whose walls are irregular interfaces. We
investigate through direct computational simulations the effect of the
interface geometry on the heat exchange efficiency of the system for
different diffusive(conductive)-convective conditions. Compared to the
behavior of a smooth channel under conditions where the convective
mechanism of transport is relevant, our results show that the activity
of the interface can be significantly underestimated if geometrical
details are not adequately considered.

This paper is organized as follows. In Section~2, we introduce the
general description of the physical system and study the flow through
the irregular geometry. The transport of heat through the flowing
fluid in the rough channel is then numerically treated in Section~3.
Finally, general conclusions and perspectives are drawn in Section~4.

\section{Viscous flow between rough interfaces}

As shown in Fig.~1, our physical system is a two-dimensional channel
of length $L$ whose delimiting walls are identical interfaces with
arbitrary geometry separated by a distance $2h$. By definition, the
two interfaces are made by the connection of several wall subsets. In
the particular roughness model adopted here, we consider that these
subsets are identical pre-fractal curves with the geometry of a
square Koch tree \cite{Evertsz92}.

The mathematical description for the detailed fluid mechanics in 
the channel is based on the assumptions that we have a continuum,
Newtonian and incompressible fluid flowing under steady state
conditions. Thus, the Navier-Stokes and continuity equations reduce to
\begin{eqnarray} 
\rho\left[u{\partial u \over \partial x} + v {\partial u\over \partial y} 
\right]&=& -{\partial p\over \partial x} 
+\mu\left[{\partial^2u\over\partial x^2} + {\partial^2u\over \partial 
y^2}\right]~, \label{eq1.a} \\ 
\rho\left[u{\partial v \over \partial x} + v {\partial v \over \partial 
y}\right] &=& -{\partial p\over \partial y} +\mu\left[{\partial^2v 
\over\partial 
x^2} + {\partial^2v \over \partial y^2}\right]~, \label{eq1.b}\\ 
{\partial u\over \partial x} + {\partial v \over \partial y}& =& 0~. 
\label{eq1.c} 
\end{eqnarray} 
Here the independent variables $x$ and $y$ denote the position in the
channel, $u$ and $v$ are the components of the velocity vector in the
$x$ and $y$ directions, respectively, and $p$ is the local pressure in
the system. The relevant physical properties of the flowing fluid are
the density $\rho$ and the viscosity $\mu$. In our simulations, we
consider non-slip boundary conditions at the entire solid-fluid
interface. In addition, the changes in velocity rates are assumed to
be zero at the exit $x=L$ (gradientless boundary conditions), whereas a
parabolic velocity profile is imposed at the inlet of the channel,
\begin{equation} 
u(0,y) = \frac{3}{2} V [1-{(\frac{y}{h})}^{2}]~. \label{eq_inletx}
\end{equation}
The Reynolds number is defined here as $Re\equiv{\rho V h / \mu}$,
where $V$ is the average velocity at the inlet. For simplicity, we
restrict our study to the case where the Reynolds number is
sufficiently low ($Re \approx 1$) to ensure a laminar viscous regime
for fluid flow. Finally, we also assume that the fluid density and
viscosity are constant properties of the flow throughout the
channel. As a consequence, the flow field is decoupled from
temperature and can be independently calculated.

The numerical solution of Eqs.~(\ref{eq1.a})-(\ref{eq1.c}) for the
velocity and pressure fields in the rough channel is obtained through
discretization by means of the control volume finite-difference
technique \cite{Patankar80}. For the complex geometry involved, we
solve this problem using a structured mesh based on quadrangular grid
elements. For example, in the case of the channel with walls subsets
that are 3-generation Koch curves, a total of 79,200 cells generates
satisfactory results when compared with numerical meshes of small
resolution. The integral form of the governing equations is then
considered at each cell of the numerical grid to produce a set of
algebraic equations which are pseudo-linearized and solved using the
SIMPLE algorithm \cite{Patankar80}. The criteria for convergence used
in the simulations is defined in terms of residuals, i.e., a measure
of the degree to which the conservation equations are satisfied
throughout the flow field. In all simulations we performed,
convergence is considered to be achieved only when each of the
residuals fall below $10^{-6}$.

In Fig.~2 we show the velocity profiles corresponding to two different
sections of the rough channel along its longitudinal
direction. Although very small when compared to the velocities at the
center, the velocities of the fluid cells constituting the reentrant
zones are still finite.  Indeed, as depicted in Fig.~3, a close-up of
these roughness details shows fluid layers in the form of consecutive
eddies whose intensities fall off in geometric progression
\cite{Moffat64,Leneweit99}. Although much less intense than the
mainstream flow, these recirculating structures are located deeper in
the system, and therefore experience closer the landscape of the
solid-fluid interface. More precisely, viscous momentum is transmitted
laterally from the mainstream flow and across successive laminae of
fluid to induce vortices inside the fractal cavity. These vortices
will then generate other vortices of smaller sizes and intensities.
In the next section, we show how this flow structure affects the
overall transport of heat as well as the distribution of fluxes 
among the active wall elements in the rough channel.

\section{Heat transport}

Once the velocity and pressure fields are obtained for the flow in the
rough channel, we proceed with the mathematical modeling and numerical
computation of the temperature scalar field. As already mentioned, we
assume that the flow is independent of the temperature. In practice,
we simply ``freeze'' the velocity field solution previously calculated
and use it to compute the steady-state temperature field $T(x,y)$
solving the two-dimensional diffusion-convection equation,
\begin{equation} \label{eq_energy} 
u\frac{\partial T}{\partial x} + v\frac{\partial T}{\partial y} =
\alpha \left( \frac{\partial^2 T}{\partial x^2} + \frac{\partial^2
T}{\partial y^2} \right)~,
\end{equation}
where $\alpha \equiv k/\rho c_{p}$ is the thermal diffusivity, $k$ 
is the thermal conductivity, and $c_{p}$ is the specific heat of the
fluid. For boundary conditions, we consider that the inlet is
subjected to a constant temperature $T_{0}$ and that the temperature
$T_{w}$ is imposed on both wall interfaces. In particular, we study
the case in which $T_{0} > T_{w}$, i.e., the heat is transported
from the bulk of the fluid to the walls of the rough channel. The
relative influence of the conductive (diffusive) and convective
mechanisms of heat transport on the behavior of the system is
quantified here in terms of the {\it P\'eclet number}, defined 
as 
\begin{equation}\label{Peclet_mac} 
Pe \equiv V h^{2} /\alpha L~.
\end{equation}
Similarly to the procedure already utilized for the calculation of the
flow, the temperature $T(x,y)$ is also obtained through discretization
of Eq.~(\ref{eq_energy}) by means of finite-differences, with an
upwind scheme to avoid numerical instabilities due to the presence 
of convection \cite{Patankar80}. Using this numerical technique,
simulations have been performed for three types of channels
corresponding to the first three pre-fractal generations of 
the square Koch tree.

In Fig.~4 we show contour plots of the temperature field in a
3-generation channel for four different diffusion-convection
conditions. At very low $Pe$ (Fig.~4a), conduction is the dominant
mechanism with the transport of heat being practically limited to the
entrance region of the channel. By increasing the value of $Pe$
(Figs.~4b-d), the contribution of convection to heat transfer
through the center of the channel becomes relevant. As shown in
Fig.~4c, although the inlet temperature can reach the exit at $Pe
\approx 200$, the temperature at the reentrant zones of the wall
remains virtually unaffected by convection. Only at very high values
of $Pe$, when the vortices inside these irregularities can effectively
transport heat, the high temperature front can penetrate deeper in the
system (Fig.~4e).

The role of the wall roughness on the transport of heat in the channel
can be quantified in terms of the normalized heat flux 
\begin{equation} \label{flux_norm} 
\phi \equiv q_{w}/q_0~,
\end{equation}
where $q_{w}$ is the total heat flux crossing the fluid-solid interface
from the bulk,
\begin{equation} \label{flux_rough} 
q_{w}=-k \displaystyle \int_{surf}~\frac{\partial T} {\partial n}~ds~.
\end{equation}
The reference value $q_0$ corresponds to the heat flux crossing a
smooth interface of size $L$ at temperature $T_{w}$ that is part of a
system where heat is transported by conduction through a static fluid
from a line at a fixed temperature $T_{0}$,
\begin{equation} \label{flux_static} 
q_{0}=k \frac{T_{0}-T_{w}}{h} L~.
\end{equation}
The dependence of the flux $\phi$ on the $Pe$ number is shown in
Fig.~5 for the first three generations of the selected pre-fractal
geometry. The behavior of a smooth channel of size $L$ and subjected
to the same boundary conditions is also shown for comparison. In this
simpler case, the system displays a sharp increase of the flux with
$Pe$, up to a point ($Pe \approx 10$) where the following power-law
regime persists for more than 5 orders of magnitude:
\begin{equation} \label{power}
\phi \sim Pe^{1/3}~. 
\end{equation}
This behavior has been observed in previous theoretical studies
\cite{Sellars56}. In the case of the power-law, it has been
demonstrated that the analytical solution of the classical Graetz
problem, where the mechanism of axial conduction is disregarded,
converges asymptotically to this scaling behavior \cite{Sellars56}. 
Alternatively, as we show in the Appendix, this typical power-law
regime of heat transport, known to be valid for moderate and high $Pe$
values, can be readily recovered by means of a global heat balance on
the system combined with a simple scaling argument.

Remarkably, the results shown in Fig.~5 indicate that different rough
channels display approximately the same behavior for low and moderate
$Pe$ numbers ($Pe < 200$). Moreover, their performances are also
similar to the behavior of the smooth channel for the same range of
$Pe$ values. Under these conditions, the conduction of heat either
predominates over convection or still represents a significant
contribution to the overall flux across the walls. 

For P\'eclet values in the range $10 < Pe < 200$, each pre-fractal
subset of the interface with the surrounding fluid can be
approximately taken as a purely diffusive cell subjected to a constant
temperature $T_{w}$ at the wall, and a temperature $T_{0}$ at a line
in the bulk of the fluid separated by a given distance from the center
of the channel.  Considering steady-state conditions, the system can
then be treated as a two-dimensional Laplacian transport unit
subjected to Dirichlet's boundary condition. This is a typical
situation where the the theorem of Makarov holds \cite{Makarov85}.
This theorem states that {\it the information dimension of the
harmonic measure on a singly connected interface in $d=2$ is exactly
equal to 1}, where $d$ represents dimension. Makarov's theorem has a
simple but important consequence: the screening effect due to
geometrical irregularities of the interface can be accessed in terms
of the ratio $S \equiv L_{p}/L$ \cite{Sapoval94}, where $L_{p}$ is the
perimeter of the interface. Because the active length of the interface
$L_{a}$ is of the order of the system size, $L_{a} \approx L$, then
$L_{a} \approx L_{p}/S$ and the factor $S$ can be considered as the
``screening factor'' of the Dirichlet-Laplacian field. When applied to
our physical system, the practical meaning of this exact result is
that, irrespective of the surface morphology, the length of the zone
within the heat exchange wall which receives most of the heat flux
should be of the order of the system size $L$, instead of the
perimeter $L_{p}$. This explains why in Fig.~5 the heat flux at the
interface is nearly insensitive to differences in the roughness
geometry for low and moderate values of $Pe$.

For $Pe > 200$, the predominance of convection over conduction is
extended to the reentrant zones of the pre-fractal units. This
crossover value results from the fact that the magnitudes of the
velocity vectors constituting the largest vortices in the rough region
are approximately two orders of magnitude smaller than the main flow
velocity $V$. At high $Pe$ values, heat can then be convected through
these recirculating structures of the flow towards the less accessible
wall elements of the interface. Under such circumstances, the theorem
of Makarov is not valid anymore, and the flux $\phi$ across the wall
becomes dependent on the geometrical details of the interface. As
shown in Fig.~5, the higher the generation of the pre-fractal
interface, larger is the heat flux at high P\'eclet conditions.

We now show that the previous analysis based on the overall heat flux
$\phi$ is consistent with the behavior of a more detailed measure of
transport. In particular, the flux heterogeneity at the interface can
be quantified in terms of an active length defined as
\cite{Sapoval99}
\begin{equation}
L_{a}\equiv 1/\sum_{i=1}^{L_{p}}\phi_{i}^2~~~~
(1 \leq L_{a} \leq L_{p})~,
\label{eq1}
\end{equation}
where the sum is over the total number of interface elements $L_{p}$,
$\phi_{i} \equiv q_{i}/\sum q_{j}$, and $q_{i}$ is the thermal flux at
the wall element $i$. From the definition (\ref{eq1}), $L_{a}=L_{p}$
indicates a limiting state of equal partition of normalized fluxes
(${\phi_i}=1/L_{p}$, $\forall i$) whereas $L_{a}=1$ should correspond
to the case of a fully ``localized'' flux distribution. The active
length therefore provides an useful index to quantify the interplay
between the Laplacian phenomenology and the complex geometry of the
absorbing interface at the local scale. As expected, the results shown
in Fig.~6 indicate that $L_{a}$ generally increases with $Pe$. It is
important to note, however, that the activities of smooth and rough
interfaces behave approximately in the same way for low values of
$Pe$. This is also consistent with Makarov's theorem under conditions
where the conduction mechanism controls the transport of heat inside
the pre-fractal roughness. 

For high P\'eclet values, $L_{a}$ becomes strongly sensitive to the
geometrical details of the interface. Again, due to the relevant
contribution of convection to the transport at the smaller scales, the
flux distribution among the interface units becomes more uniform. The
higher the pre-fractal generation used to create the roughness
geometry, higher is the available number of wall elements for heat
exchange. As a consequence, a substantial increase in the active
length can be observed for a fixed value of $Pe$.

\section{Conclusions}

The development of a coherent scientific approach to the project of
heat and mass transport devices involving new concepts from
mathematics and physics still represents an important research and
technological challenge to chemical and mechanical engineers.  In the
present work, we have studied the conditions under which the roughness
geometry of a channel with a flowing fluid should play a relevant role
on the transport of heat. For low P\'eclet numbers, when the mechanism
of heat conduction dominates, it is shown that the overall flux of
heat across a pre-fractal interface subjected to Dirichlet boundary
conditions has little sensitivity to details and irregularities of
the heat absorption zone. Under similar conditions, the length of the
active zone computed for a smooth channel is very close to the
measure obtained for a rough system. This behavior is then interpreted
in terms of a screening phenomenon that privileges the absorbing
activity of the geometrically exposed sites of the interface.
Moreover, such a ``localization'' of the heat flux can be formally
explained in terms of Makarov's theorem for Laplacian transport in
two-dimensional irregular geometries.

The situation is entirely different at high P\'eclet numbers. In this
case, our simulations reveal that both the heat flux and the active
length start to depend noticeably on the morphological details of the
interface. In this situation, not only the total heat flux leaving the
system increases faster for higher generations of the pre-fractal
interface, but the distribution of fluxes at the wall units also
becomes gradually less ``localized'' ($L_{a}$ increases) as $Pe$
increases. This transition reflects the onset of convective effects on
the heat transport in the proximity of the rough interface. A closer
look at the system indicates that the significant changes observed in
$L_{a}$ above the transition point are induced by the presence of
recirculation zones of flow (vortices) in these regions.

As a future work, we will extend the approach introduced here to
simulate the effect of different geometries on the transport
efficiency of the system as well as other types of ``absorption''
mechanisms (e.g., finite-rate heat transfer at the walls) limited by
diffusion transport. Finally, we also believe that the results
presented in this study provide useful information to the
understanding of other physico-chemical systems with relevance to
science and technology that include heat, mass or momentum
transport. The modeling strategy devised here should also be helpful
as a design tool to choose a suitable interface geometry for a given
heat exchange system.

\section{Acknowledgments}

We thank Bernard Sapoval and Marcel Filoche for many helpful
discussions. We acknowledge the Brazilian agencies CNPq, CAPES and
FUNCAP for financial support.

\section{Appendix: The smooth channel}

Here we revisit the classical problem of heat transport through a
fluid flowing in a channel whose walls are smooth interfaces. The
system setup is shown in Fig.~7. As in the case of the rough channel,
we consider a Newtonian and incompressible fluid with constant
properties that is flowing under viscous and steady-state
conditions. Assuming non-slip boundary conditions at the walls, the
velocity along the channel follows the classical parabolic profile
\begin{equation}\label{perfil_vel}
u = \frac{3}{2}V(1-{\eta}^2)~,
\end{equation}
where $\eta \equiv y/h$, $V \equiv \Delta p~h²/(3 \mu L)$, and
$\Delta p$ is the pressure drop between the entrance and the exit of
the channel. Under these simple flow conditions, the fluid is suddenly
subjected to the temperature $T_{0}$ at the channel inlet, while a
constant temperature $T_w$ is imposed at both walls. At steady-state,
and assuming that the axial heat conduction mechanism is negligible,
this corresponds to the so-called {\it Graetz problem}, where the
temperature field in the channel obeys the following differential
equation:
\begin{equation}\label{eq_energy2} 
u(y) \frac{\partial T}{\partial x} = \alpha \frac{\partial^2
T}{\partial y^2}~,
\end{equation}
with boundary conditions
\begin{eqnarray}\label{bc_energy} 
T(0,y)=T_{0}~~~~~\text{and}~~~~~T(x,h)=T_{w}~.
\end{eqnarray}
An analytical solution for the linear partial differential equation
(\ref{eq_energy2}) subjected to (\ref{bc_energy}) can be obtained
through separation of variables and a proper expansion in terms of the
so-called {\it Graetz functions} \cite{Arpaci66,Sellars56}. From this
solution, it is possible to show that the overall heat flux through
the absorbing walls, $\phi$, follows the power law behavior, $\phi
\sim Pe^{1/3}$, that is valid for moderate and high $Pe$ values.

The objective here is to show that this power-law regime of heat
transport can be readily recovered by means of a global heat balance
combined with a simple scaling argument. First, in the absence of
sources or sinks of heat, the total heat crossing the walls should be
equal to the difference
\begin{equation}\label{heat_balance} 
q_w = q_{in}-q_{out}~,
\end{equation}
where $q_{in}$ and $q_{out}$ are the fluxes at the entrance and at the
exit sections of the channel, respectively. At sufficiently high
values of the P\'eclet number, the convective part of the flux
dominates the heat transfer process. As shown in Fig.~8, the
temperature profile at the exit displays a central plateau in the
range $-y^*\le y\le y^*$, where the temperature is equal to the
entrance temperature $T_0$, and then decays monotonously towards the
wall temperature $T_w$, in the region $y^*\le |y| \le h$.  The value
of $\eta^*(\equiv y^*/h)$ can be found from the definition of the {\it
local P\'eclet number}
\begin{equation}\label{peclet} 
Pe_{\eta} \equiv \frac{\tau_D}{\tau_C}=\frac{\delta^2/\alpha}{L/u(\eta)}~,
\end{equation}
where $\tau_D \equiv \delta^2/\alpha$ is the characteristic conduction
time, $\tau_C \equiv L/u$ is the local convective time, and $\delta
\equiv h(1-\eta^*)$. If we now argue that $\eta^*$ should correspond
to the position where $Pe_{\eta} \approx 1$, we can write from
Eqs.~(\ref{perfil_vel}) and (\ref{peclet}) that
\begin{equation}
\frac{3}{2} \frac{V h^2}{\alpha L} (1-\eta^*)^2[1-(\eta^*)^2]=1~,
\nonumber
\end{equation}
which, from the definition (\ref{Peclet_mac}), leads to
\begin{equation}
(1-\eta^*)^3(1+\eta^*)=\frac{2}{3Pe}~.
\nonumber
\end{equation}
For high values of $Pe$, $\eta^*$ is very close to 1, such that
$(1-\eta^*)^3 (1+\eta^*) \approx 2(1-\eta^*)^3$, resulting in
\begin{equation}\label{approximation}
\eta^*\approx 1 - \frac{1}{(3Pe)^{1/3}}~.
\end{equation}
Going back to Eq.~(\ref{heat_balance}), the heat flux $q_{in}$
entering the system is given by
\begin{equation}
q_{in} = 2\rho c_p  T_0  \int^h_0 u(y)\,dy~.
\nonumber
\end{equation}
The flux $q_{out}$ leaving the channel can be expressed as the sum
\begin{equation}
q_{out}=q_{1}+q_{2}~,
\end{equation}
where the first term corresponds to the heat escaping through the
plateau region of temperature $T_0$ and size $y^*$
\begin{equation}
q_{1} = 2\rho c_p  T_0 \int ^{y^*}_0 u(y)\, dy~,
\end{equation}
while the second term represents the flux through the remaining part
of the channel cross-section with variable temperature
\begin{equation}\label{q_2}
q_{2}=2\rho c_p \int^h_{y^*}  u(y)T(y)\,dy~.
\end{equation}
As a first approximation, if we assume that the temperature in this
region decreases linearly with $y$,
\begin{equation}\label{T}
T(y)=\frac{T_0-T_w}{y^*-h}(y-y^*)+T_0~,
\nonumber
\end{equation}
Eq.~(\ref{q_2}) can then be written as
\begin{eqnarray}
q_{2}&=&2\rho c_p T_0 \int^h_{y^*} u(y) \, dy \nonumber \\
&+& 2\rho c_p \int^h_{y^*} u(y)\frac{T_0-T_w}{y^*-h}(y-y^*)\, dy~,
\nonumber  
\end{eqnarray}
and the total flux becomes
\begin{equation}\label{total_y}
q_w=2\rho c_p \frac{T_{0}-T_w}{h-y^*} \int_{y^*}^h
\frac{3}{2}V\left[1- \left( \frac{y}{h}\right)^2\right]
(y-y^*)dy~.
\nonumber
\end{equation}
Rewriting this expression in terms of the dimensionless variable
$\eta$, we obtain
\begin{eqnarray}
q_w &=& 3 \rho c_p (T_{0}-T_w) V h \frac{1}{1-\eta^*}
\int_{\eta^*}^1 \left(1-\eta^2\right)
(\eta-\eta^*)d\eta \nonumber \\
&=& \frac{3}{12} \rho c_p (T_{0}-T_w) V h
\left[(\eta^*)^3 + (\eta^*)^2- 5\eta^* +3\right]~.
\nonumber
\end{eqnarray}
From this equation and the definitions (\ref{Peclet_mac}) and
(\ref{flux_norm}), the normalized flux can be expressed as
\begin{equation} \label{polynomial}
\phi = \frac{3}{12} Pe \left[(\eta^*)^3 + (\eta^*)^2-5 \eta^* +3\right]~,
\nonumber
\end{equation}
and using the approximation (\ref{approximation}), we finally get the
scaling relation
\begin{equation} \label{scaling}
\phi \approx \frac{\left(3 Pe \right)^{1/3}}{3} -\frac{1}{12}~.
\end{equation}

\begin{figure}
\includegraphics[width=8cm,angle=-90]{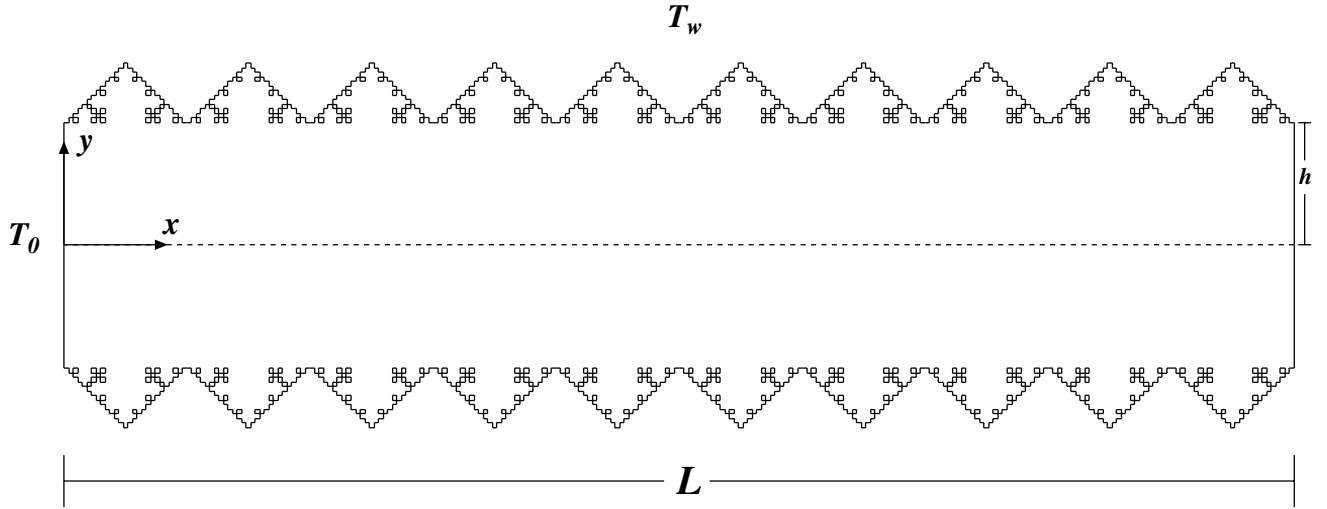}
\vspace{1.5cm}
\caption{Schematic diagram of the rough channel under study.
The walls are a sequence of irregular interfaces with the geometry of
square Koch trees. The fluid flows steadily from left to right at $Re
\approx 1$. A constant temperature $T_{0}$ is maintained at the inlet,
while the temperature on both walls is $T_{w}$. We study the
steady-state transport of heat for the case in which $T_{0} > T_{w}$,
i.e., the heat goes from the bulk of the fluid to the walls of the
channel.}
\end{figure}

\begin{figure}
\includegraphics[width=10.0cm,angle=-90]{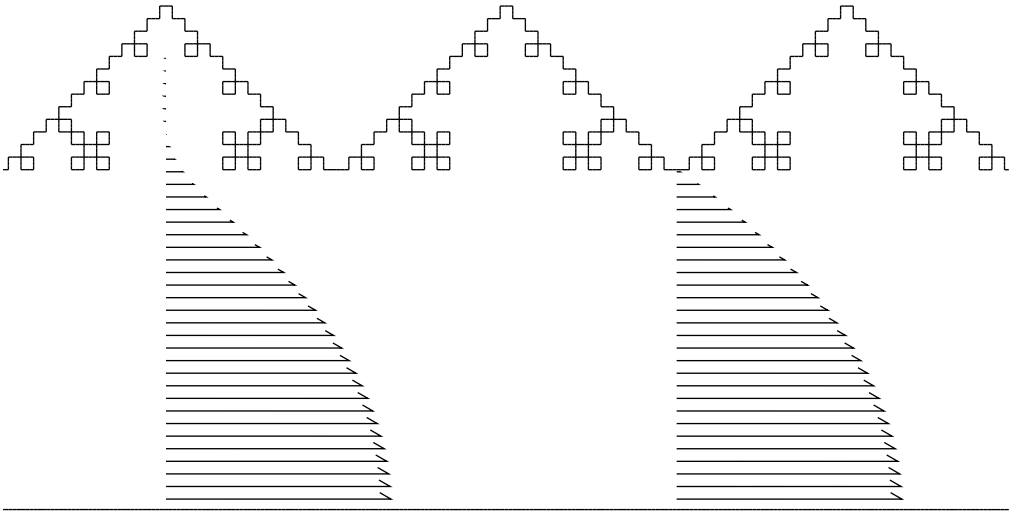}
\caption{Longitudinal velocity profiles at two distinct cross-sections
of the rough channel. Although small when compared to the mainstream
flow, the velocities inside the reentrant zones of the pre-fractal
geometry are not equal to zero, as can be observed in the profile on
the left. At the cross-section on the right, the velocity profile is
typically parabolic.}
\end{figure}

\begin{figure}
\includegraphics[width=10.0cm]{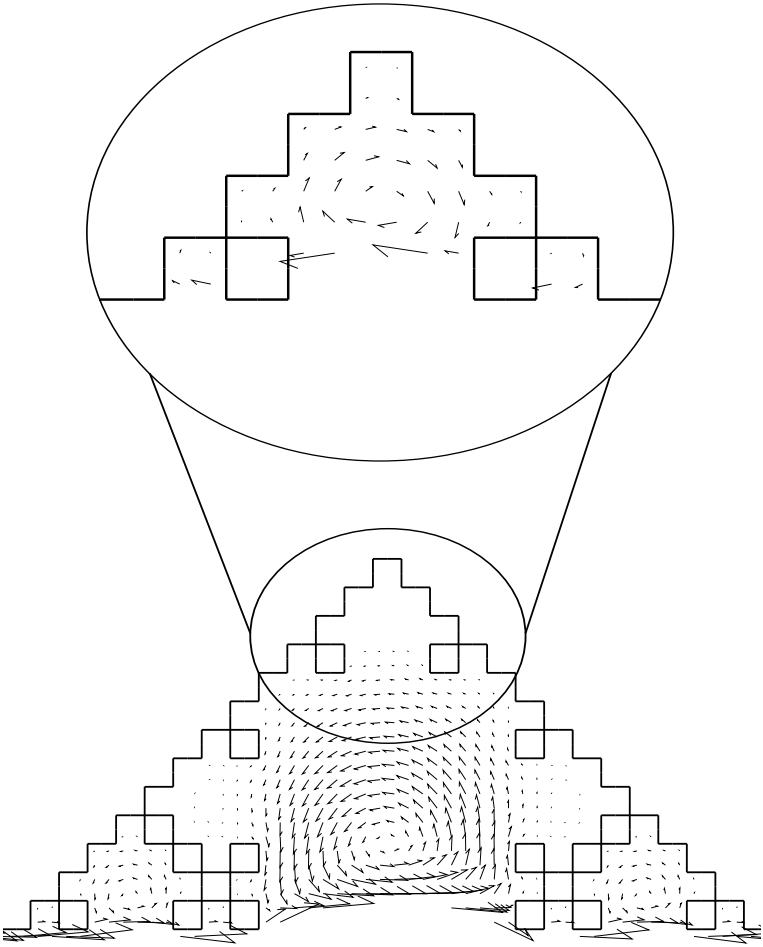}
\caption{Vortices in the reentrant zones of the pre-fractal roughness.
A close-up view of a smaller roughness detail reveals the sequence of
eddies with intensities that fall off in geometric progression.}
\end{figure}

\begin{figure}
\includegraphics[width=10.0cm]{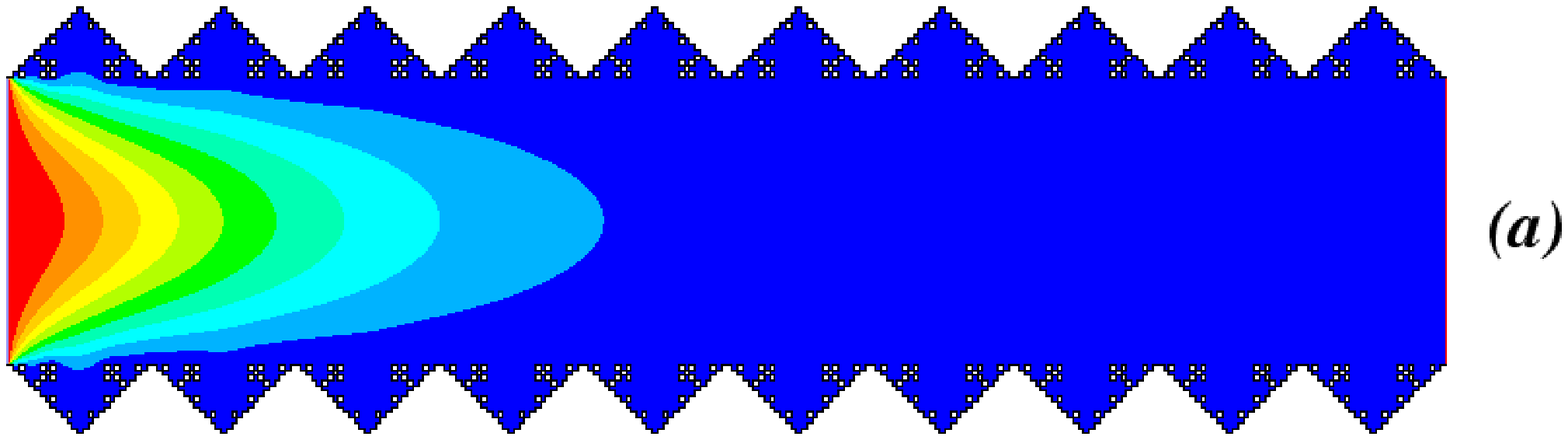}
\includegraphics[width=10.0cm]{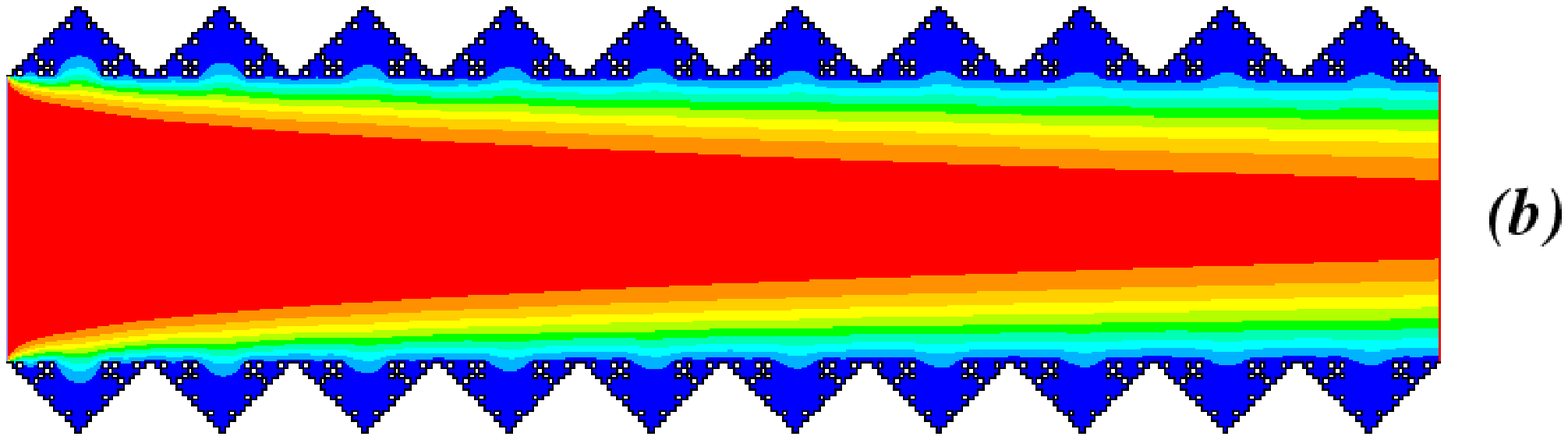}
\includegraphics[width=10.0cm]{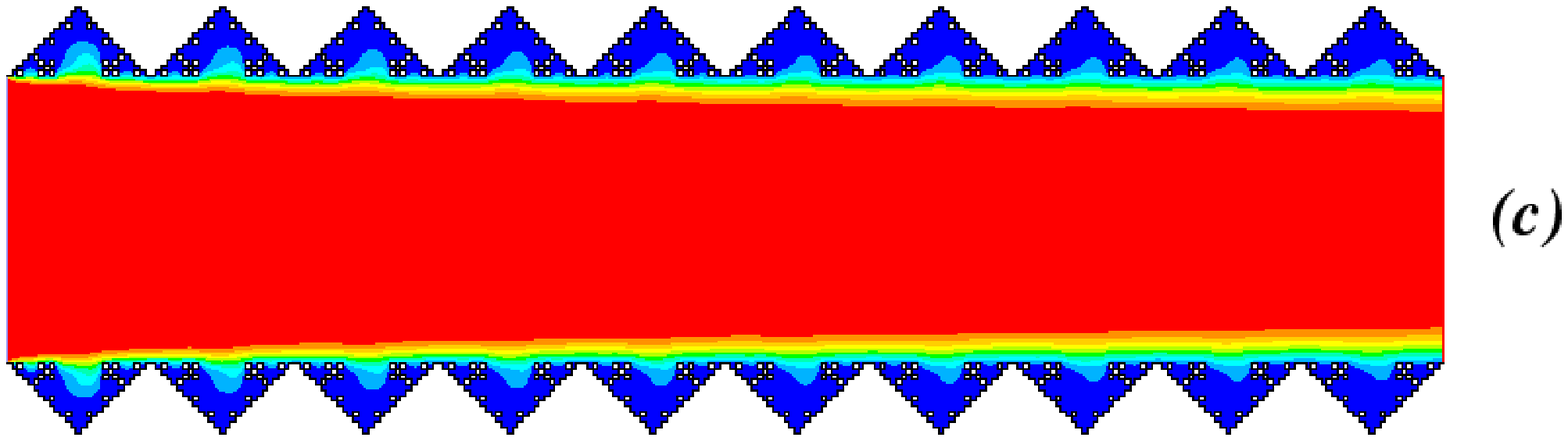}
\includegraphics[width=10.0cm]{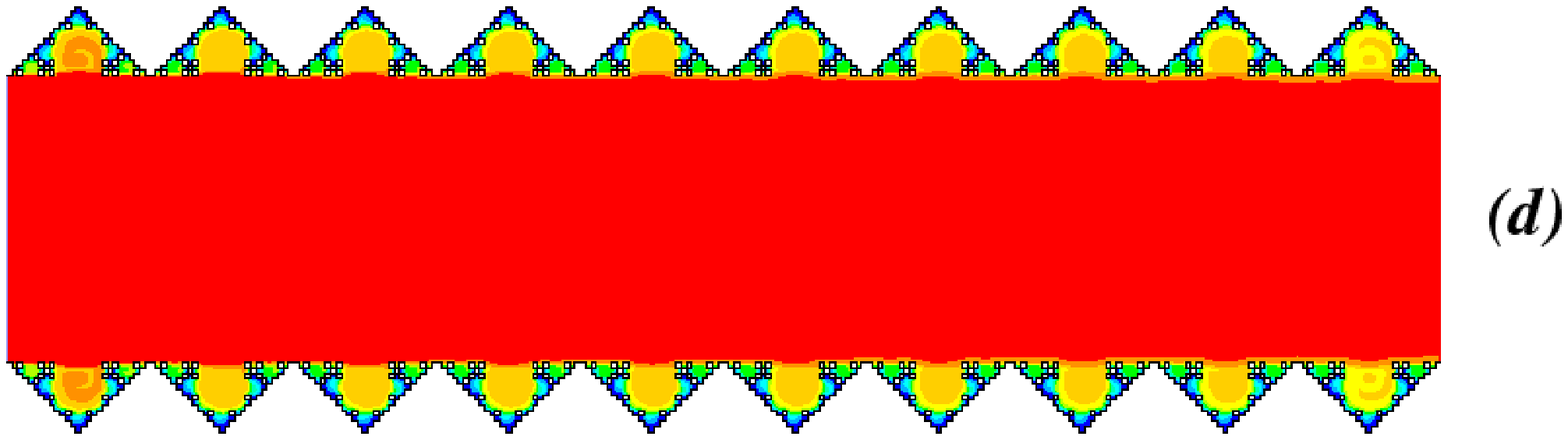}
\includegraphics[width=10.0cm]{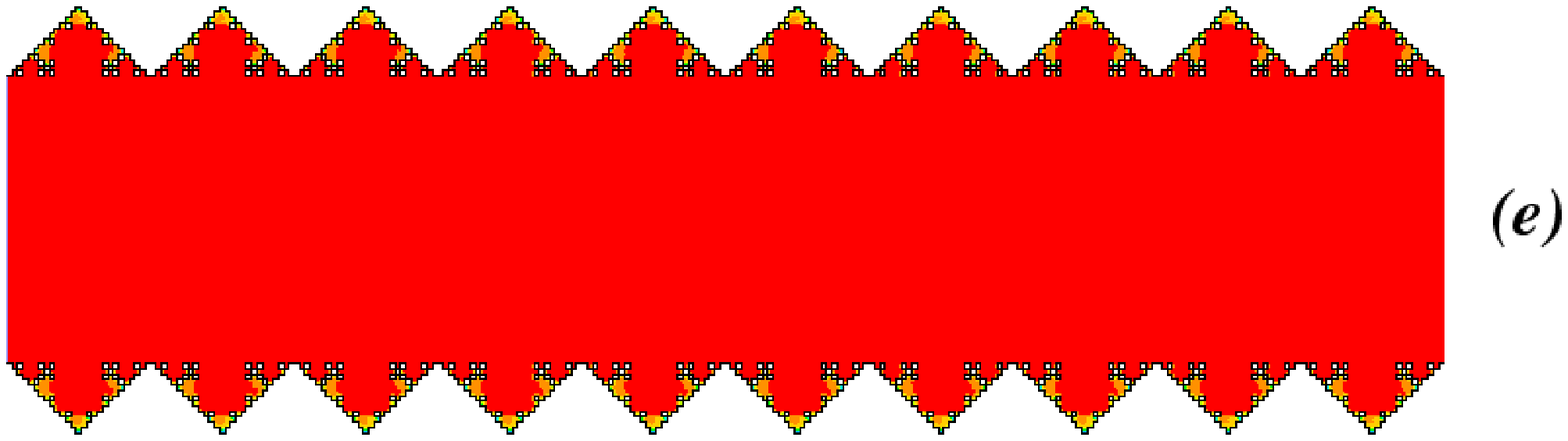}
\caption{Contour plot of the temperature field for different values
of the P\'eclet number: (a) $Pe=0.25$, (b) $Pe=10$, (c) $Pe=200$,
(d) $Pe=10^5$, and (e) $Pe=10^7$. The temperature decreases from
red (gray dark) to blue (gray light).}
\end{figure}

\begin{figure}
\includegraphics[width=10.0cm]{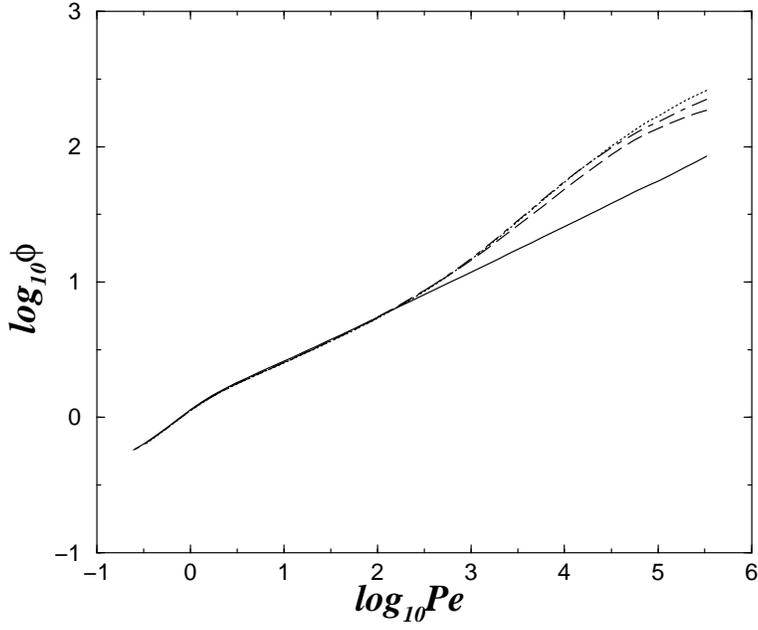}
\caption{Log-log plot showing the dependence of the normalized
flux across the walls, $\phi$, on the P\'eclet number, $Pe$. The solid
line represents the behavior of a smooth channel, while the dashed,
dot-dashed and dotted lines correspond to the first, second and third
pre-fractal generations, respectively.}
\end{figure}

\begin{figure}
\includegraphics[width=10.0cm]{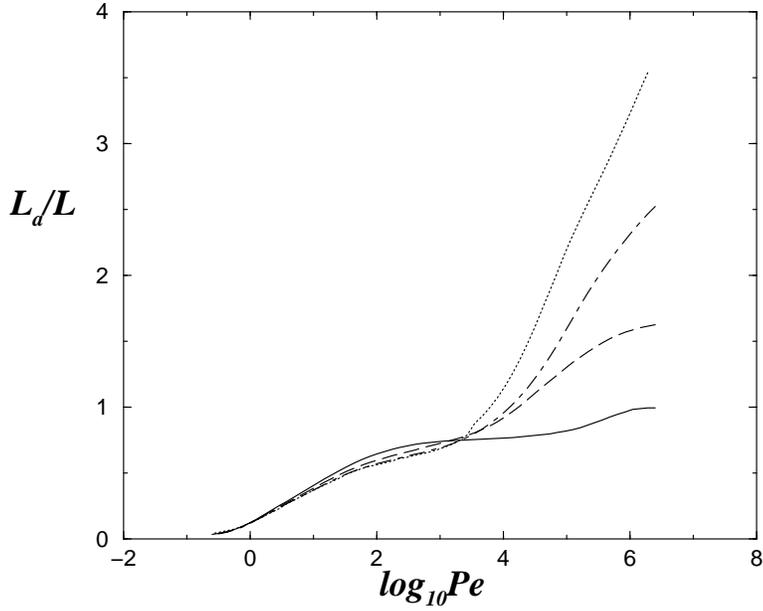}
\caption{Logarithmic plot showing the variation of the length of the
active zone, $L_{a}$, with the P\'eclet number, $Pe$. The solid line
corresponds to the case of a smooth channel, while the dashed,
dot-dashed and dotted lines represent the behavior of the first,
second and third pre-fractal generations, respectively.}
\end{figure}

\begin{figure}
\includegraphics[width=10.0cm]{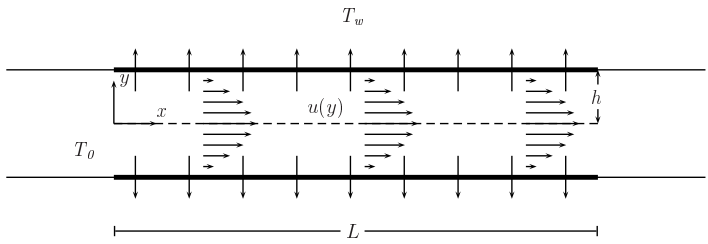}
\vspace{0.2cm}
\caption{Schematic diagram of the smooth channel showing the
fully developed parabolic profile of the flow. As in the case of the
rough channel, the inlet is at temperature $T_{0}$ while the
temperature on both walls is maintained at $T_{w}$. We also consider
the case in which $T_{0} > T_{w}$.}
\end{figure}

\begin{figure}
\includegraphics[width=10.0cm]{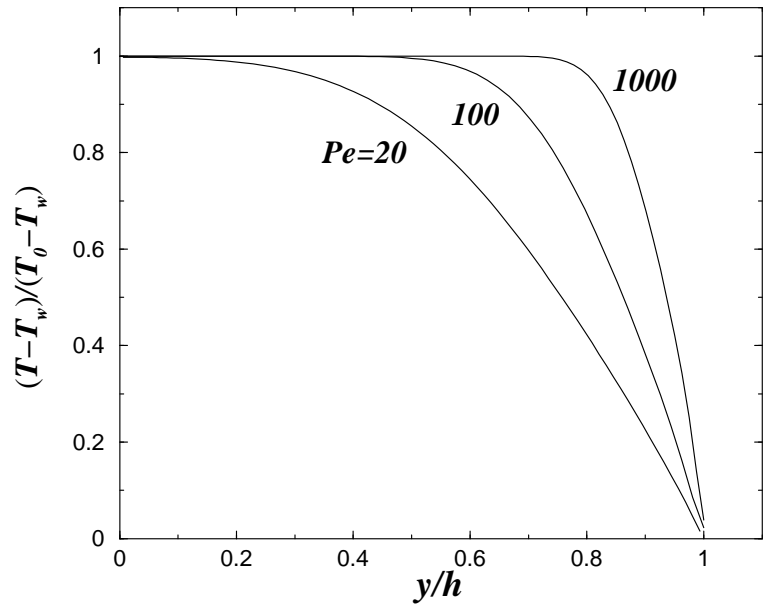}
\caption{Steady-state profiles of the normalized temperature at the 
exit of the smooth channel for three different values of the P\'eclet
number.}
\end{figure}

\end{document}